# A novel approach to muscle functional recovery monitoring by its free vibration measurement

Agnieszka Tomaszewska[1][*], Milena Drozdowska[1], Piotr Aschenbrenner [2]

[1] Gdansk University of Technology, Faculty of Civil and Environmental Engineering, Department of Structural Mechanics, Narutowicza 11/12, 80-233 Gdansk, Poland
[2] Gdansk University of Physical Education and Sport, Department of Biomechanics and Sports Engineering, Kazimierza Gorskiego 1, 80-336 Gdansk,  Poland

**Abstract**

The study shows a potential to monitor functional recovery of a muscle by its natural frequency measurement.  A novel approach is presented, in which vibrations of the muscle of interest are measured in a contactless manner using laser displacement sensor and the measured vibration signals are analyzed by an advanced technique of experimental modal analysis to extract fundamental natural frequency of the muscle with high accuracy. A case study of functional status recovery of a rectus femoris muscle, which become athropic after ACL reconstruction, is presented. Three time points of the patient's recovery process are considered: a day before the patient's going back to normal sport activity after ACL reconstruction, one month later and one year later. The analysis shows significant changes of the fundamental natural frequency of the rectus femoris within the muscle recovery process, the frequency approaches the value identified in the same muscle in the other, reference leg of the patient. The proposed approach is worth application in a wider group of patients. The method displays a practical potential to monitor the rehabilitation progress, to assess muscular functional status of dysfunctional patients or to assess muscular readiness attitude to undertake sport competition.

## 1.  Introduction

Vibration pattern of a solid provides an information on its technical condition. That is known from the times of Chinese pottery invention, which quality was tested by striking it gently and listening to its sound response (Rytter, 1993). Today, the knowledge around vibration-based structural health monitoring (SHM) is abundant. One group of SHM schemes is based on modal characteristics analysis. Such an approach identifies modal parameters, i.e. natural frequencies, mode shapes and damping coefficients based on vibration signals. The modal parameters are controlled by system data, i.e. mass, stiffness, tension state (e.g. prestress), thus they indicate system evolution. They are not affected by external actions. The modal parameters analysis in SHM is advantageous - the data required for their identification can be measured non-invasively, e.g. during regular system operation. Numerous modal identification techniques are tailored to specific sorts of excitation and structural response. Various techniques and their application in civil and mechanical engineering is addressed in books (Ewins, 2000), (Rainieri and Fabbrocino, 2014).

---

[*] corresponding author, email address: atomas@pg.edu.pl (Agnieszka Tomaszewska)

The SHM finds its application in a field of biomechanics too. Vibration measurement of human body members may be conducted non-invasively, to yield biomechanical properties of a given body part or to analyse a specific bodily behaviour. As an example, vibration-based analysis of stiffness and tension of Achilles tendon is considered in (Salman and Sabra, 2015) and (Martin et al., 2018). Full-field measurement of facial vibrations, using electronic speckle pattern interferometry is addressed in (Frič et al., 2018). The latter authors study the relation between the facial vibrations and singing.

In the study a SHM methodology is applied to a biomechanical assessment problem of a muscle functional status by its fundamental natural (resonant) frequency identification. The inquiry is focused on rectus femoris, which often becomes atrophic after anterior cruciate ligament (ACL) reconstruction (ACLR) in the knee joint. The rupture of ACL is frequent in human population. Various sources denote 30 to 78 ACL ruptures per 100,000 person-years (Gans et al., 2018). Moreover, it is hard to restore pre-injury strength of a quadriceps, even despite intensive rehabilitation (Mattacola et al., 2015), (Ithurburn et al., 2018).

There are various means available to assess muscular functional status, e.g. by measuring strength of muscles acting on the knee joint using a Biodex system (Mattacola et al., 2015), or muscle tone or local stiffness measured using MyotonPRO (Aird et al., 2012), or shear wave elastography (SWE) to measure tissue elasticity (Nowicki and Dobruch-Sobczak, 2016). However a number of drawbacks occur here. The Biodex system does not recognize a single muscle. The Myoton measures muscle vibrations in contact with the tissue, which impacts the measurement results, due to gentle tissue vibrations. Finally, the SWE equipment is expensive, it requires an expertise operator.

The article shows a novel approach to muscle functional recovery monitoring. Free muscular vibrations are measured in a contactless mode, the fundamental natural frequency is identified by advanced experimental modal analysis technique. The assumption of pre-injury limbs symmetry makes it possible to monitor the rehabilitation progress of the atrophic rectus femoris muscle by comparison of its fundamental frequency with frequency of the healthy muscle, in another leg of the same patient. The concept is discussed in the following case study.

## 2. Case Presentation

A single female patient case is considered here. The patient signed an informed consent to the research. She is 45 years old, of 18.6 BMI, sedentary work, sport amateur practicing dance (2x2h weekly) and horse riding (2x1h weekly). She ruptured the ACL in her right leg (ACL leg) while skiing. She underwent ACLR, right after the quadriceps in the ACL leg became atrophic. One month after the reconstruction rehabilitation started, with laser and magnetic field therapy, and unloading exercises of the knee joint. It took two weeks, which is a Polish standard. The therapy included riding a stationary bicycle every day, several quarter sessions. Two months after ACLR the patient started to run short distances. Five months after ACLR the patient was able to run and ride a standard bicycle. And here is the first time point of the analysis (August 2022), in which first vibration measurement was taken. A second time point, to bring subsequent measurement, came after a month, in September 2022. The period between the first and second time points saw the patient return to horse riding, 3 times a week, 1h each ride. The third time point was dated after a year, in September 2023. The year between the second and the third time points was the patient's time to implement pre-injury sport activities.

The analysis concerns the process monitoring of a functional status of rectus femoris muscle retrieval in the ACL leg related to the muscle in the other, healthy (reference) leg, in the same patient. The three time points were considered, every time muscular free vibrations were measured by a laser displacement sensor optoNCDT 1402 ILD1402-600 (Micro-Epsilon, Germany), of 80μm resolution and the measurement rate set to 100 Hz. After preliminary measurements and signals analysis at different points of the muscles in both legs, the point at the highest rectus femoris muscle diameter, in front of the patient was selected for vibration tracking. Displacement time series were measured in the sagittal plane, in horizontal direction. An outlook of the measurement stand is presented in Fig. 1a. The muscle vibrations were induced by a single hit in the upper region. The hitting point was covered by a small coin, to prevent from hematoma formation due to test repetition. The hit was made by a modal hammer (model 086D05, PCB piezotronics), force applied to the muscle was measured. The load was applied at least seven times in each tested case.

Two states of both considered muscles (rectus femoris in both legs) were included in the study, i.e. the relaxed and tense (voluntary maximal contraction) ones. The patient was located in standing position, subjectively symmetrical. A selected force signal applied to rectus femoris in ACL leg in a tense state at the second time point is shown in Fig.1b, accompanied by relevant muscular responses (Figure 1b,c).

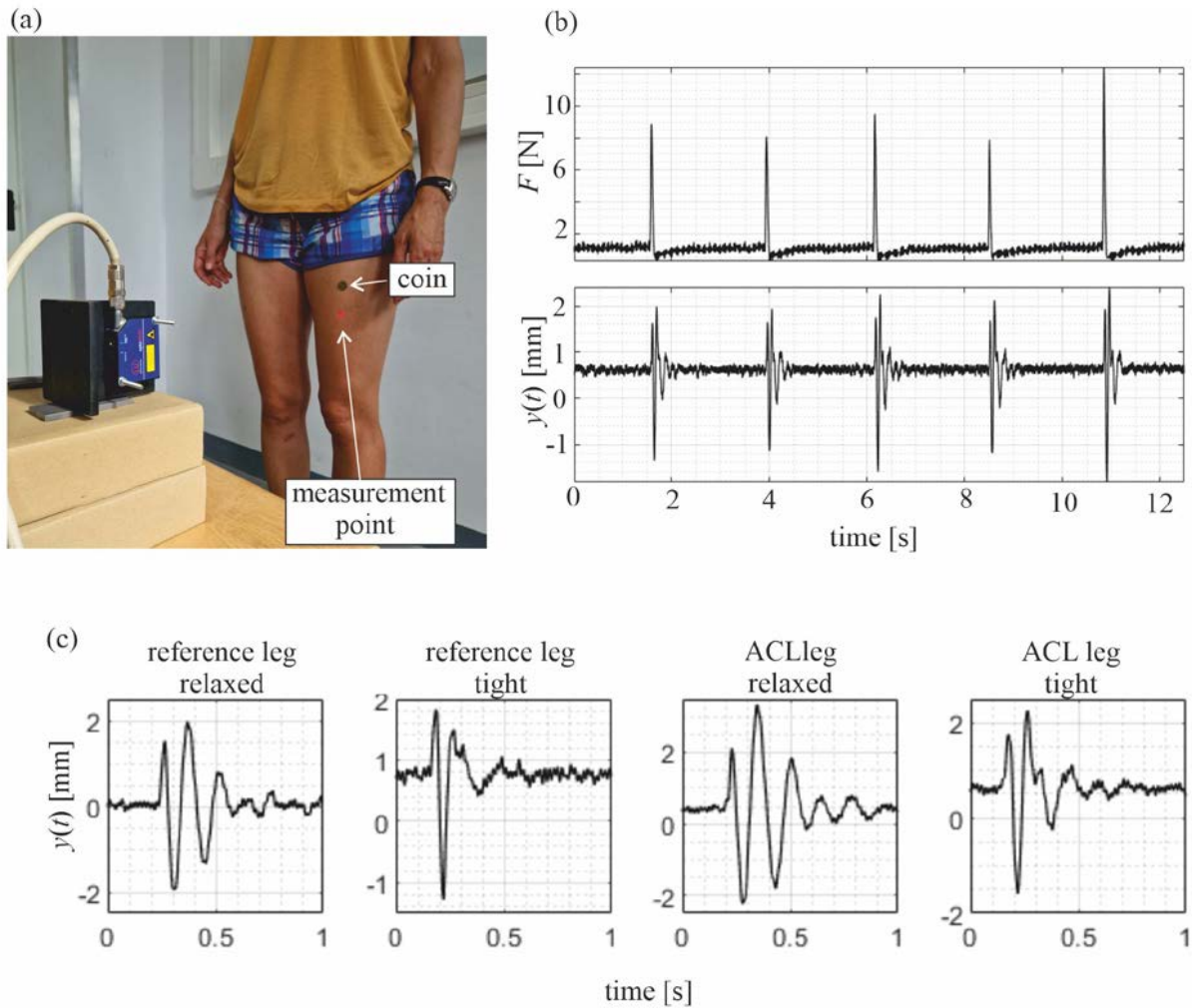

Fig. 1 (a) View of the experimental stand; (b) Signals measured in ACL leg – force *F* and vibration response *y* of the rectus femoris muscle in a tense state (second time point); (c) Insight on muscular responses to a single force input (second time point).

### 3. Identification of the muscle natural frequency from free vibration signals

Natural muscular frequencies were extracted from the response signals by an Eigensystem Realization Algorithm (ERA) (Juang and Pappa, 1985), designed to analyze free vibrations of a system of decaying character, which resembles the study case. The technique is uniquely effective in identifying dynamic characteristics of systems, such as natural frequencies, mode shapes and damping ratios. It is popular in mechanical and civil engineering (Tomaszewska and Szafrański, 2020). High identification accuracy is triggered by a stabilization diagram related to different orders of the analysis, it rejects the bias error. The method is significantly superior to Fourier Transform used sometimes to identify natural frequencies, which however delivers a general information only on frequencies contributing the analyzed signal.

The ERA is based on estimation of the state-space representation of a linear dynamic system. The state space model of a discrete-time system is reflected by a system of equations

$$\begin{cases} \dot{\mathbf{x}}(t) = \mathbf{A}_c \mathbf{x}(t) + \mathbf{B}_c \mathbf{u}(t) \\ \mathbf{y}(t) = \mathbf{C}_c \mathbf{x}(t) + \mathbf{D}_c \mathbf{u}(t) \end{cases}, \quad (1)$$

where **A**, **B**, **C** and **D** are state, input, output and transmission matrices respectively, **u**(*t*), **x**(*t*) and **y**(*t*) are input, state and output vectors, respectively. Parts of a time-domain decaying signal (triggered by an impulsive input) form a Hankel matrix of Markov parameters $y_k$. Considering a single output **y**(*t*) the matrix includes subsequent signal samples at discrete time points *k*.

$$\mathbf{H}(k-1) = \begin{bmatrix} y_k & y_{k+1} & \cdots & y_{k+\beta-1} \\ y_{k+1} & y_{k+2} & \cdots & y_{k+\beta} \\ \vdots & \vdots & \ddots & \vdots \\ y_{k+\alpha-1} & y_{k+\alpha} & \cdots & y_{k+\alpha+\beta-2} \end{bmatrix} \in \mathbb{R}^{\alpha \times \beta}, \quad (2)$$

where $\alpha$, $\beta$ are parameters to determine the size of Hankel matrix. Next, Singular Value Decomposition (SVD) is performed on the Hankel matrix, to bring three matrices **S**, **U**, **V**: the matrix made of singular values of the Hankel matrix, arranged in a non-increasing order and two non-singular orthonormal matrices, made of vectors corresponding to singular values of **S** matrix, respectively

$$\text{SVD}\mathbf{H}(0) = \mathbf{U}\mathbf{S}\mathbf{V}^T. \quad (3)$$

The most significant part of the **S** matrix is selected to specify the model order *n*. This is the most important part of the ERA procedure. While the model order is too low the important modal identification results may be neglected. The order too high may lead to false modes identification due to signal noise.

$$\mathbf{S}_{\alpha m \times \beta} = \begin{bmatrix} \mathbf{S}_n & \mathbf{0} \\ \mathbf{0} & \hat{\mathbf{S}} \\ \mathbf{0} & \mathbf{0} \end{bmatrix}, \quad \text{where} \quad \mathbf{S}_n = \begin{bmatrix} \sigma_1 & 0 & 0 & 0 \\ 0 & \sigma_2 & 0 & 0 \\ 0 & 0 & \ddots & 0 \\ 0 & 0 & 0 & \sigma_n \end{bmatrix}, \quad \hat{\mathbf{S}} = \begin{bmatrix} \hat{\sigma}_{n+1} & 0 & 0 & 0 \\ 0 & \hat{\sigma}_{n+2} & 0 & 0 \\ 0 & 0 & \ddots & 0 \\ 0 & 0 & 0 & \hat{\sigma}_\beta \end{bmatrix}. \quad (4)$$

Taking the first *n* columns of **U** and **V** and an $n \times n$ part of **S**, the Hankel matrix is reconstructed respective to the selected model order

$$\widetilde{\mathbf{H}}(k-1)_{(\alpha \times \beta)} = \mathbf{U}_{n \, (\alpha \times n)} \mathbf{S}_{n \, (n \times n)} \mathbf{V}_{n \, (n \times \beta)}^T, \quad \mathbf{U}_{(\alpha \times \alpha)} = \begin{bmatrix} \mathbf{U}_n & \hat{\mathbf{U}} \end{bmatrix}, \quad \mathbf{V}_{(\beta \times \beta)} = \begin{bmatrix} \mathbf{V}_n & \hat{\mathbf{V}} \end{bmatrix}. \quad (5)$$

These computations are provided at two time steps, $k=1$ to achieve $\mathbf{U}_n, \mathbf{S}_n, \mathbf{V}_n$ matrices and $k=2$ to estimate the state matrix $\hat{\mathbf{A}}$

$$\hat{\mathbf{A}} = \mathbf{S}_n^{-1/2} \mathbf{U}_n^T \widetilde{\mathbf{H}}(1) \mathbf{V}_n \mathbf{S}_n^{-1/2}. \quad (6)$$

Finally, solution of the eigenvalue problem of the state matrix identifies the modal parameters

$$\hat{\mathbf{A}}\boldsymbol{\Phi} = \boldsymbol{\Phi}\hat{\boldsymbol{\Lambda}}. \quad (7)$$

Matrix $\widehat{\mathbf{\Lambda}}$ is diagonal, it includes pairs of conjugate eigenvalues

$$\widehat{\mathbf{\Lambda}} = \mathrm{diag}\begin{bmatrix} \hat{\lambda}_1 & \hat{\lambda}_1^* & \hat{\lambda}_2 & \hat{\lambda}_2^* & \cdots & \hat{\lambda}_n & \hat{\lambda}_n^* \end{bmatrix}. \qquad (8)$$

Natural frequencies $f_m$ are related with the eigenvalues, here $f_s$ is the data sampling frequency

$$\widehat{\mathbf{\Lambda}}_c = f_s \ln(\widehat{\mathbf{\Lambda}}), \qquad \omega_m = |\hat{\lambda}_c|, \qquad f_m = \frac{\omega_m}{2\pi}. \qquad (9)$$

## 4. Results and discussion

The analytical results of the collected vibration signals are presented in Table 1 and in Figure 2. The indicator of the functional status of rectus femoris in the ACL leg is an absolute ($\Delta f$) or relative ($\Delta f/f_r$) difference between the fundamental frequency identified for this muscle and a twin muscle in a reference leg, with regard to a given time point and a given muscular state (relaxed or tense).

Table 1. Natural frequencies of rectus femoris in ACL and reference legs

| Muscular state | Time point | reference leg | | ACL leg | | $\Delta f = f_r - f_a$ [Hz] | $\Delta f / f_r$ |
|---|---|---|---|---|---|---|---|
| | | $f_r$ [Hz] | standard deviation | $f_a$ [Hz] | standard deviation | | |
| relaxed | VIII.2022 | 11.01 | 1.55 | 9.06 | 1.22 | 1.95 | 0.177 |
| | IX.2022 | 7.74 | 0.29 | 7.06 | 0.09 | 0.68 | 0.088 |
| | IX.2023 | 8.75 | 0.74 | 8.29 | 0.26 | 0.46 | 0.053 |
| tense | VIII.2022 | 15.57 | 1.78 | 10.88 | 0.50 | 4.69 | 0.301 |
| | IX.2022 | 14.54 | 0.64 | 13.06 | 0.43 | 1.48 | 0.102 |
| | IX.2023 | 11.37 | 1.62 | 12.07 | 1.39 | -0.70 | -0.062 |

Table 2. Mean values of peak torque measured for reference ($T_r$) and ACL ($T_a$) legs in two time points

| Time point | $T_r$ [Nm] | $T_a$ [Nm] | $\Delta T = T_r - T_a$ [Nm] | $\Delta T / T_r$ |
|---|---|---|---|---|
| IX.2022 | 94 | 69 | 25 | 0.266 |
| IX.2023 | 95 | 101 | -6 | -0.063 |

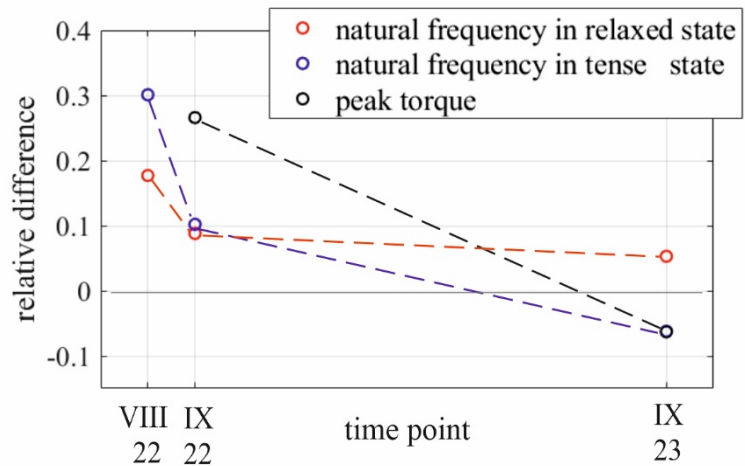

Fig. 2. Relative differences in natural frequencies of rectus femoris muscle and in peak torque in knee joint in reference and ACL leg

As a reference, knee joint torque was measured in both legs at the 2$^{nd}$ and 3$^{rd}$ time points by a Biodex System 4 PRO (Biodex Medical System Inc., Shirley, NY, USA). Isometric unilateral extension with 90 degree bend was used, preserving a 30s relaxation time and 5s contraction time. Three repetitions were made each time. The acquired mean values are presented in Table 2 and a relative difference between reference and ACL legs is plotted in Figure 2.

The analysis reveals various fundamental natural frequencies of the considered muscles, i.e. rectus femoris in a reference and ACL legs, in both considered tension states. At the first time point the difference in values obtained for both legs is extremely visible in the muscular tense state. Here relative difference between natural frequencies is 0.301 (see Table 1 and Figure 2). In the relaxed state a smaller, but still considerable variation equals 0.177. At subsequent time points the variation decreased - the third time shows the values of -0.062 and 0.053 in the tense and relaxed states of the muscles, respectively. Variation between natural frequencies of the rectus femoris muscles in the reference and ACL legs proves that initially, at the first time point the two considered muscles were in fact distinct structures. It may be concluded on the basis of the relaxed state analysis alone, however, the tense state analysis makes the conclusion firm. It means that the discrepancy in natural frequencies between both considered muscles becomes even greater than in the relaxed state, which means rectus femoris in the ACL leg does not yield tension close to the level in the reference leg muscle. This thesis is confirmed by torque measurement in knee joints in both considered legs. At the second time point the relative peak torque variation in both knees is 0.266, it subsequently drops to -0.063 at the last time point.

Hence the natural frequencies point out variations in functional status in both considered muscles at first two time points and similar status of both muscles at the last time point. The conclusion on the muscle functional status is confirmed in torque measurements, so the proposed approach may be used to assess muscular functional status. This novel scheme may be interesting alternative to existing ones, especially as the measurement is contactless and noninvasive, simple, fast-made and does not demand expensive device. The results were presented at DHM Congress (Tomaszewska et al., 2023).

## 5. Conclusions

The study presents a potential to monitor functional recovery of a muscle by means of its fundamental natural frequency measurement. The proposed approach should be validated on a broad group of patients. The new approach finds application in rehabilitation progress monitoring, muscle functional status assessment in patients after some contusion or in assessment of muscular ability to undertake sport challenge.

**Credit authorship contribution statement**

**Declaration of Competing Interest**

The authors declare that they have no known competing financial interests or personal relationships that could have appeared to influence the work reported in this paper.

**Acknowledgment**